\documentclass[usenatbib]{basi}
\usepackage[british]{babel}
\usepackage[varg]{txfonts}
\usepackage{rotating}
\usepackage{dcolumn}

\newcommand{\phx}{\emph{PHOENIX}}
\newcommand{\muse}{\emph{MUSE}}

\begin{document}
\title[A New Library of Synthetic Spectra from PHOENIX and its Application to Fitting MUSE Spectra]{A New Extensive Library of Synthetic Stellar Spectra from PHOENIX Atmospheres and its Application to Fitting VLT MUSE Spectra}
\author[T.-O.~Husser et~al.]%
       {T.-O.~Husser$^1$\thanks{email: \texttt{husser@astro.physuk.uni-goettingen.de}}, S.~Kamann$^2$, S.~Dreizler$^1$, Peter.~H.~Hauschildt$^3$\\
       $^1$Institut f\"ur Astrophysik, Georg-August-Universit\"at G\"ottingen, Friedrich-Hund-Platz 1, \\ \quad 37077 G\"ottingen, Germany\\
       $^2$Leibniz-Institut f\"ur Astrophysik Potsdam (AIP), An der Sternwarte 16, 14482 Potsdam, Germany\\
       $^3$Hamburger Sternwarte, Gojenbergsweg 112, 21029 Hamburg, Germany}

\pubyear{2012}
\volume{00}
\pagerange{\pageref{firstpage}--\pageref{lastpage}}

\date{Received \today}

\maketitle
\label{firstpage}

\begin{abstract}
We present a new library of synthetic spectra based on the stellar atmosphere code \phx. It
covers the wavelength range from 500\,\AA\ to 55\,000\,\AA\ with a resolution of R=500\,000 in the
optical and near IR, R=100\,000 in the IR and $\Delta\lambda=$0.1\,\AA\ in the UV. The parameter space covers
$2\,300\,\mathrm{K} \le T_{\mathrm{eff}} \le 8\,000\,\mathrm{K}$, $0.0 \le \log(g) \le +6.0$, $-4.0 \le [Fe/H] \le +1.0$
and $-0.3 \le [\alpha/Fe] \le +0.8$.
The library is work-in-progress and going to be extended to at least $T_{\mathrm{eff}}=25\,000\,\mathrm{K}$. We use
a new self-consistent way of describing the microturbulence for our model atmospheres.
The entire library of synthetic spectra will be available for download.\\
Futhermore we present a method for fitting spectra, especially designed to work with the new
2nd generation VLT instrument \muse. We show that we can determine stellar parameters ( $T_{\mathrm{eff}}$,
$\log(g)$, $[Fe/H]$ and $[\alpha/Fe]$) and even single element abundances. 
\end{abstract}


\section{Introduction}
Presumably next year the Multi Unit Spectroscopic Explorer \muse\ \citep{2010SPIE.7735E...7B} will see its first light
as a second-generation instrument for \emph{ESO}'s \emph{Very Large Telescope} at Paranal, Chile. The instrument is an highly efficient AO-supported integral field unit (IFU)
and its outstanding combination of a large field of view ($1 \times 1\,\mathrm{arcmin}^2$) and high spatial sampling
of 0.2" with a spectral resolution of R=2000-4000 over the optical range of 4650-9300\,\AA\
will allow us to obtain unprecedented observations.

In our group, we intent to use it for the analysis of galactic globular clusters, which due to the heavy
crowding towards the center are only accessible through their giants by other instruments. With \muse,
we will be able to point directly in the center of the clusters and obtain thousands of stellar spectra 
even from stars well below the main-sequence turnoff point with one single exposure.

For the analysis of those spectra, we need a grid of model spectra that matches both the wavelength range
and resolution of \muse\ as well as our requirements for an extensive parameter space (as given by
previous observations of globular clusters) and being able to adjust it as needed. Therefore a decision
was made to create a new grid of model atmospheres and synthetic spectra with \phx\ \citep{1999JCoAM.109...41H}.

We will present this new library together with a short description of the methods used for
analyzing \muse\ spectra and some preliminary results on a simulated data cube.
A paper \citep{husser2012} about our new synthetic stellar library with more detailed descriptions
and informations for downloading the spectra is in preparation and will be published
soon.

\section{Globular Clusters}
There are about 150 known globular clusters in our galaxy with masses of $10^5-10^6\,M_\mathrm{sun}$, which consist
of very old stellar populations with an age of $\geq 10\,\mathrm{Gyr}$. A couple of years ago those populations were
assumed to be very simple with a single isochrone.

Unexpectedly, latest observations showed evidence for the existence of multiple main-se\-quen\-ces within globular
clusters, e.\,g.\ \cite{2004ApJ...605L.125B} for Omega Centauri, and also for a split 
in the (sub)giant branch \citep{1999Natur.402...55L}. This seems to be caused by multiple stellar populations with different abundances of
helium and iron, at least for massive clusters. A variation 
of some lighter elements seems to be more ubiquitos, like the Na-O anti-correlation observed by Carretta (2009). Those
observations indicate a complex enrichment history with multiple epochs of star formation.

In globular clusters, we observe a velocity dispersion of $5-20\,\mathrm{km/s}$ and high central stellar densities
of $\sim$10$^6\,\mathrm{M}_\mathrm{sun}/\mathrm{pc}^3$. Given these high stellar densities, scenarios have been proposed 
for the formation of intermediate-mass black holes in the cluster centres \citep{2002ApJ...576..899P}. An extrapolation of the tight relation between total mass and the 
mass of the black hole in the bulges of galaxies (Marconi and Hunt 2003) towards typical masses of globular clusters predicts black
holes with a mass of $\sim$10$^3\,\mathrm{M}_\mathrm{sun}$.

In contrast to field stars, where we observe a binary fraction of about 50\%, in globular clusters we
find a value of only 30\% or even less. Monte Carlo simulations \citep{2005MNRAS.358..572I} showed that
the number of binaries in the core decreases rapidly over time. \cite{1987degc.book.....S} already showed
that a depletion of binaries in the core is necessary for it to collapse, so core-collapsed clusters
like M15 seem to be dynamically more evolved than others. A core-collapse could also result in the formation
of new binaries. Therefore, studying the binary fraction in globular clusters is an important task in order to understand their evolution.

\section{The new \phx\ grid}
The \phx\ version 16 that we are using for calculating the grid uses a new equation of state called ACES,
which is a state-of-the-art treatment of the chemical equilibrium in each layer of a stellar atmosphere.
The element abundances we used for the atmospheres were taken from \cite{2009ARA&A..47..481A}.

\renewcommand{\tabcolsep}{0.5mm}
\begin{table}
  \caption{Parameter space of the grid. An extension in $T_{\mathrm{eff}}$ up to $12\,000\,\mathrm{K}$ is work in progress and 
	   up to $25\,000\,\mathrm{K}$ in planning.
	   Alpha element abundances $[\alpha/Fe] \neq 0$ are only available for $3\,500\,\mathrm{K} \leq T_{\mathrm{eff}} \leq 8\,000\,\mathrm{K}$ 
	   and $-2 \leq [Fe/H] \leq 0$.}
  \label{table:paramspace}
  \centering                                      %
  \begin{tabular}{rrrlc}
    \hline \hline
			   & & \multicolumn{2}{c}{Range}  & Step size \\ \hline
    $T_{\mathrm{eff}}$ [K] & & 2\,300 & -- 7\,000              & 100 \\
			   & & 7\,000 & -- 8\,000              & 200 \\
    $\log(g)$         	   & &   0.0 & -- +6.0     & 0.5 \\
    $[Fe/H]$               & &  -4.0 & -- -2.0    & 1.0 \\
		           & &  -2.0 & -- +1.0    & 0.5 \\
    $[\alpha/Fe]$          & &  -0.3 & -- +0.8     & 0.1 \\ \hline
  \end{tabular}
\end{table}
\renewcommand{\tabcolsep}{2mm}
The parameter space of the new \phx\ grid that we are presenting in this paper is given in Table~\ref{table:paramspace}.
An extension towards hotter stars including NLTE treatment of important elements is both work-in-progress (up to 12\,000\,K) and intended 
(up to 25\,000\,K). The grid is complete in its first three dimensions effective temperature $T_{\mathrm{eff}}$, surface 
gravity $\log(g)$ and metallicity $[Fe/H]$. Different alpha element (including O, Ne, Mg, Si, S, Ar, Ca and Ti) abundances are provided for 
$3\,500\,\mathrm{K} \leq T_{\mathrm{eff}} \leq 8\,000\,\mathrm{K}$  and $-2 \leq [Fe/H] \leq 0$ only.

\renewcommand{\tabcolsep}{0.5mm}
\begin{table}
  \caption{Spectral resolution of the grid.}
  \label{table:resolution}
  \centering                                      %
  \begin{tabular}{rlrl}
    \hline \hline
    \multicolumn{2}{c}{Range [\AA]}      & \multicolumn{2}{c}{Resolution} \\ \hline
        500 & -- 3\,000    & $\Delta\lambda$ & = $0.1$\AA \\
     3\,000 & -- 25\,000   & $R$ & $\approx 500\,000$ \\
    25\,000 & -- 55\,000   & $R$ & $\approx 100\,000$ \\ \hline
  \end{tabular}
\end{table}
\renewcommand{\tabcolsep}{2mm}
Although the primary intent for creating the grid was the analysis of \muse\ spectra, we decided to increase both the
wavelength range and the resolution (see Table~\ref{table:resolution}), so that teams working on other existing and upcoming 
instruments will be able to use them for their purposes. Due to to this, our spectra are applicable for the analysis of
e.\,g.\ CRIRES \citep{2004SPIE.5492.1218K} and X-Shooter \citep{2011A&A...536A.105V} data.

\begin{figure}
  \includegraphics[width=\textwidth]{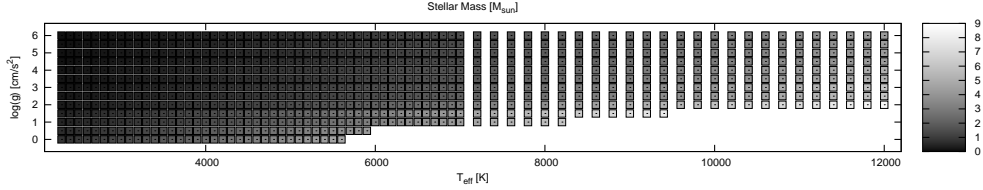}
  \caption{Distribution of stellar masses for that part of the grid with solar abundances for different
           effective temperatures $T_{\mathrm{eff}}$ and surface gavities $\log(g)$. Color-coded is the
	   stellar mass in units of solar mass from $0M_\odot$ (black) to $9M_\odot$ (white).}
  \label{figure:mass}
\end{figure}
In order to define a spherical symmetric atmosphere as it is used in \phx, we need to define an 
effective temperature $T_{\mathrm{eff}}$, a surface gravity $\log(g)$ and either a radius $r_0$ or a mass
$M_\star$. We decided to use the mass by taking a mass-luminosity relation $L_\star/L_\odot = (M_\star/M_\odot)^3$
for main-sequence stars and letting it tend towards higher values for giants and super giants:
\begin{equation}
  M_{\star} = c \cdot M_{\mathrm{sun}	} \cdot \left( \frac{T_{\mathrm{eff}}}{5\,770\,\mathrm{K}} \right)^2,
  \label{eq:math}
\end{equation}
with values for the coefficient as given in the following table:
\begin{center}
  \begin{tabular}{r|ccccccc}
    \hline \hline
    log(g) & $>4$  & $>3$  & $>2$  & $>1.6$  & $>0.9$  & $>0$  & $\leq0$ \\
    c      & 1     & 1.2   & 1.4   & 2       & 3       & 4     & 5 \\ \hline
  \end{tabular}
\end{center}
Figure~\ref{figure:mass} shows the distribution of masses in our grid for solar abundances. 

\phx\ uses the mixing length theory \citep{Prandtl,1953ZA.....32..135V} for describing convection within the atmosphere.
For our spectra, we used the formula provided by \citet{1999A&A...346..111L}, which has been calibrated
using 3D RHD models:
\begin{equation}
  \alpha = a_0 + (a_1 + (a_3 + a_5 T_s + a_6 g_s) T_s + a_4 g_s) T_s + a_2 g_s,
\end{equation}
with
\begin{equation}
T_s = \frac{T_{\mathrm{eff}} - 5\,770\,\mathrm{K}}{1\,000} \quad\mathrm{and}\quad g_s = \log \left( \frac{10^{\log(g)}}{27\,500} \right),
\end{equation}
and coefficients given by:
\begin{center}
  \begin{tabular}{c|c|c|c|c|c|c}
    $a_0$  & $a_1$  & $a_2$  & $a_3$  & $a_4$  & $a_5$  & $a_6$  \\ \hline
    1.587  & -0.054 & 0.045  & -0.039 & 0.176  & -0.067 & 0.107
  \end{tabular}
\end{center}

\begin{figure}
  \centering
  \includegraphics[width=\textwidth]{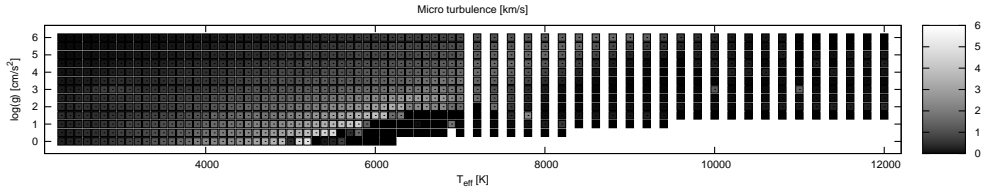}
  \caption{Distribution of micro-turbulences for that part of the grid with solar abundances for different
           effective temperatures $T_{\mathrm{eff}}$ and surface gavities $\log(g)$. Color-coded is the
	   micro-turbulence from $0\textrm{km/s}$ (black) to $6\textrm{km/s}$ (white).}
  \label{figure:microTurb}
\end{figure}
For matching synthetic spectra with observed ones, we need micro-turbulence as an additional adhoc parameter.
Our definition of this is that a large scale (macro-) turbulent motion triggers a small scale
(micro-) turbulent motion on length scales below the photon main free path length, which affects the strength of spectral lines \citep{gray2005observation}.
In this picture, the micro-turbulence is strongly related to macro-turbulent motion, therefore we use 
$v_{\mathrm micro} = 0.5 \cdot \left< v_{\mathrm conv} \right>$ as an experimental
formula that follows from 3D radiative hydrodynamic investigations of cool M-stars \citep{2009A&A...508.1429W}.
So we first calculate the model atmosphere and then synthesize a spectrum from it using a micro-turbulence,
which is assumed to be half the mean convective velocity in the photosphere. Fig.~\ref{figure:microTurb}
shows the distribution of micro-turbulences in our grid for solar abundances.
Unfortunately we had a problem with \phx\ concerning convection for giants around 7\,000\,K, so we had to disable convection
for those models. Therefore there is no micro-turbulence included as well.

\section{Fitting stellar parameters}	
The spectroscopic analysis of crowded stellar fields, such as star clusters or nearby galaxies has been limited to 
relatively small samples of stars thus far. The main problem in this respect is that traditionally used techniques 
like multi-object spectroscopy are restricted to the brighter, isolated stars in the field. We have developed a 
new method to overcome this limitation using integral field spectroscopy. Taking advantage of the combined spatial 
and spectral coverage provided by an integral field spectrograph, we developed a new analysis approach which we 
call ``crowded field spectroscopy'' \citep{2004AN....325..155B,Kamann}. Via PSF fitting techniques, single object spectra for the stars above the confusion 
limit are extracted. This deblending technique works so well that we obtain clean stellar spectra for a significantly 
higher number of stars than hitherto possible. 

For the extracted spectra we determine the stellar parameters using a weighted constrained
non-linear least-squares minimization (Levenberg-Marquardt), similar to the ULySS package by
\cite{2009A&A...501.1269K}, which has been used as well e.\,g.\ by \cite{2011RAA....11..924W}. Usually six different parameters are fitted: effective 
temperature $T_{\mathrm{eff}}$, surface gravity $\log(g)$, metallicity $[Fe/H]$, $\alpha$-element abundance $[\alpha/Fe]$,
radial velocity $v_\mathrm{rad}$ and line broadening $\sigma$.

In every iteration of the Levenberg-Marquardt algorithm, a model spectrum is extracted from the \phx\ grid
using an N-dimensional spline interpolator. Then a line of sight velocity distribution (LOSVD) is applied to the
model, which adds line broadening $\sigma$ and a shift caused by e.\,g.\ radial velocity $v_{\mathrm{rad}}$. Finally an N-dimensional Legendre
polynomial (i.\,e.\ the continuum difference between model and observation) is determined using a linear fit in a way that 
the model multiplied with this polynomial matches the observation as best as possible. Due to this we are independent 
of the continuum and the fit is done on spectral lines only.

\begin{figure}
  \centering
  \includegraphics[height=5.3cm]{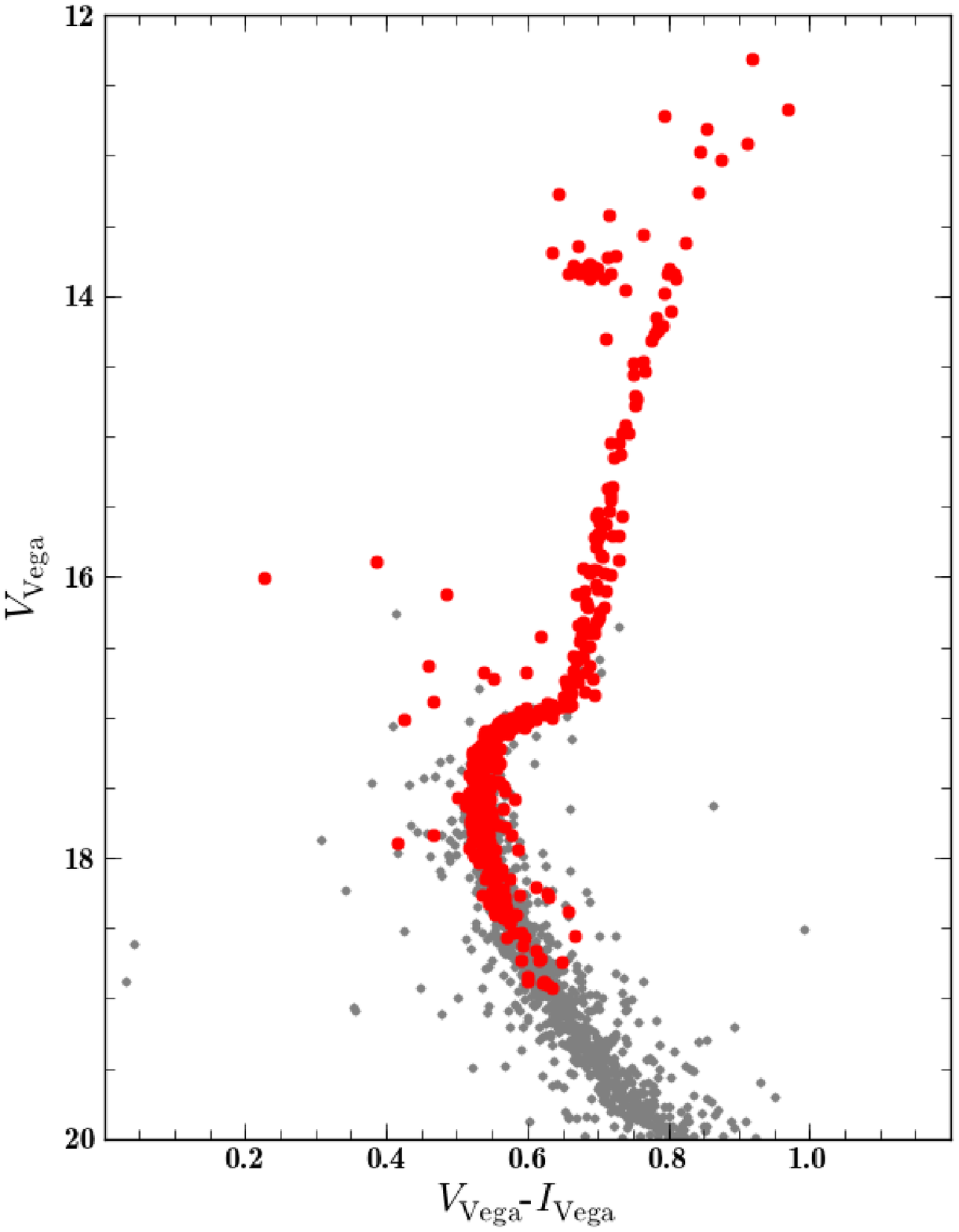} 
  \quad
  \includegraphics[height=5.3cm]{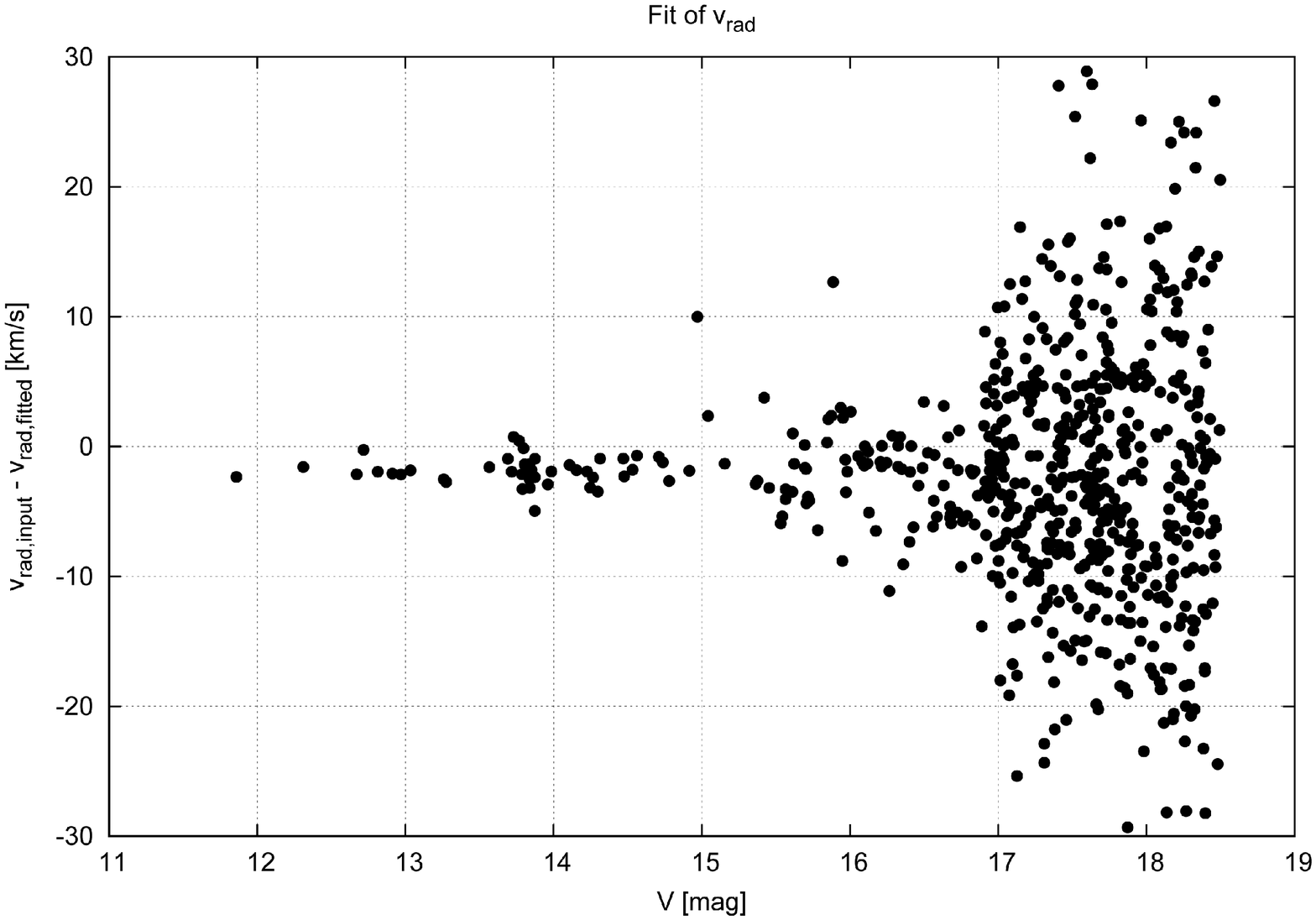} 
  \caption{The color-magnitude diagram for a simulated \muse\ data cube is shown on the left. For the
           analysis we only used 1/9 of the cube, for which the stars are marked in red. On the right the
           errors in fitted radial velocities are plotted.}
  \label{figure:dryRun}
\end{figure}
Figure~\ref{figure:dryRun} shows some first preliminary results for a simulated \muse\ data cube based
on real HST observations of 47 Tuc, obtained by \cite{2007AJ....133.1658S}. As one can see, at the main-sequence
turnoff point at $\sim$17mag we still can fit the radial velocity with an accuracy of about 5km/s. The systematic
offset in the results is caused by a known error in the creation of the simulated data cube. At that magnitude,
the corresponding error in the fitted effective temperature is of the order of 100\,K.

\cite{2008ApJ...682.1217K} showed that even with medium resolution spectra it is possible to determine the abundance of single
elements, in their case it was some of the alpha elements (Mg, Si, Ca, Ti). For this analysis, they created a mask in order to fit only those parts
of a spectrum against a grid of synthetic spectra, where it changes most when varying the analysed element. They
created the mask by looking for regions, where two spectra with a given $T_{\mathrm{eff}}$, a fixed $\log(g)$, 
a metallicity of $[Fe/H]=-1.5$ and element abundances of $[X/Fe]=\pm0.3$ differ by more then $0.5\%$. Masks for several
different temperatures where combined into a single mask that was used for the analysis.
The main advantage of this method was that they did not need new dimensions in the grid for 
every new element, but could use an existing one (here $[\alpha/Fe]$), since the masks did not overlap.

With our new \phx\ grid, we can go one step further and create a mask specifically
for every single spectrum that we want to analyse. Therefore we use the same method as described by
\cite{2008ApJ...682.1217K}, but use the previously fitted values for $T_{\mathrm{eff}}$ and $\log(g)$ of 
the observed spectrum. Using the individual mask for each star, alpha element abundances can then be
determined with an uncertainty of typically 0.1-0.2dex.
Of course our intention is to extent this method even further in order to fit the
abundance of other elements.

When observing the same field multiple times, we will have radial velocities for several epochs, so that
we can determine orbital parameters for binaries that we find. Furthermore we can extend our method described
above to fit simultaneously the two components of a binary star and henceforth derive the atmospheric 
parameters of both stars.

\section{Conclusion}
We presented a new extensive grid of synthetic stellar spectra from \phx\ atmospheres with
a wavelength range and resolution that should cover all existing and upcoming instruments. Currently
its parameter range is optimized for the analysis of globular clusters, but we intend to extend
it to higher temperatures. We also introduced a new self-consistent way of describing the
micro-turbulence in model atmospheres.

Furthermore we presented a first view on the methods for analyzing globular clusters with data
obtained with the \emph{VLT} \muse\ together with preliminary results on simulated data. We showed that we
will be able to examine both the kinematics as well as the binary fraction. In addition we will have
accurate stellar parameters for most of the stars in the field.


\label{lastpage}

\begin{thebibliography}{}
 \bibitem[Asplund et al.(2009)Asplund et al.]{2009ARA&A..47..481A} 
  Asplund M., Grevesse N., Sauval A.~J., Scott P., 2009, ARA\&A, 47, 481 
 \bibitem[Bacon et al.(2010)Bacon et al.]{2010SPIE.7735E...7B} 
  Bacon R., et al., 2010, SPIE, 7735,  
 \bibitem[Becker et al.(2004)Becker, Fabrika, \& Roth]{2004AN....325..155B} 
  Becker T., Fabrika S., Roth M.~M., 2004, AN, 325, 155 
 \bibitem[Bedin et al.(2004)Bedin et al.]{2004ApJ...605L.125B} 
  Bedin L.~R., Piotto G., Anderson J., Cassisi S., King I.~R., Momany Y., Carraro G., 2004, ApJ, 605, L125 
 \bibitem[Gray (2005)Gray]{gray2005observation}
  Gray D.~F., 2005, The observation and analysis of stellar photospheres (Cambridge University Press)
 \bibitem[Hauschildt \& Baron(1999)Hauschildt \& Baron]{1999JCoAM.109...41H} 
  Hauschildt P.~H., Baron E., 1999, JCoAM, 109, 41 
 \bibitem[Husser et al.(in prep.)Husser et al.]{husser2012}
  Husser T.-O. et al., 2012, in preparation
 \bibitem[Ivanova et al.(2005)Ivanova et al.]{2005MNRAS.358..572I} 
  Ivanova N., Belczynski K., Fregeau J.~M., Rasio F.~A., 2005, MNRAS, 358, 572 
 \bibitem[Kaeufl et al.(2004)Kaeufl et al.]{2004SPIE.5492.1218K} 
  Kaeufl H.-U., et al., 2004, SPIE, 5492, 1218 
 \bibitem[Kamann(in prep.)Kamann]{Kamann} 
  Kamann S., 2012, in preparation
 \bibitem[Kirby et al.(2008)Kirby, Guhathakurta, \& Sneden]{2008ApJ...682.1217K} 
  Kirby E.~N., Guhathakurta P., Sneden C., 2008, ApJ, 682, 1217 
 \bibitem[Koleva et al.(2009)Koleva et al.]{2009A&A...501.1269K} 
  Koleva M., Prugniel P., Bouchard A., Wu Y., 2009, A\&A, 501, 1269 
 \bibitem[Lee et al.(1999)Lee et al.]{1999Natur.402...55L} 
  Lee Y.-W., Joo J.-M., Sohn Y.-J., Rey S.-C., Lee H.-C., Walker A.~R., 1999, Natur, 402, 55 
 \bibitem[Ludwig et al.(1999)Ludwig, Freytag, \& Steffen]{1999A&A...346..111L} 
  Ludwig H.-G., Freytag B., Steffen M., 1999, A\&A, 346, 111 
 \bibitem[Mihalas (1978)Mihalas]{1978stat.book.....M}
  Mihalas D., 1978, Stellar atmospheres /2nd edition/, ed. Mihalas, D.
 \bibitem[Portegies Zwart \& McMillan(2002)Portegies Zwart \& McMillan]{2002ApJ...576..899P} 
  Portegies Zwart S.~F., McMillan S.~L.~W., 2002, ApJ, 576, 899 
 \bibitem[Prandtl (1925)Prandtl]{Prandtl}
  Prandtl L., 1925, Zeitschr. Angewandt. Math. Mech., 5, 136
 \bibitem[Sarajedini et al.(2007)Sarajedini et al.]{2007AJ....133.1658S} 
  Sarajedini A., et al., 2007, AJ, 133, 1658 
 \bibitem[Spitzer L. (1987)Spitzer]{1987degc.book.....S}
  Spitzer L., Dynamical evolution of globular clusters, ed. Spitzer, L.
 \bibitem[Vernet et al.(2011)Vernet et al.]{2011A&A...536A.105V} 
  Vernet J., et al., 2011, A\&A, 536, A105 
 \bibitem[Vitense(1953)Vitense]{1953ZA.....32..135V} 
  Vitense E., 1953, ZA, 32, 135 
 \bibitem[Wende et al.(2009)Wende, Reiners, \& Ludwig]{2009A&A...508.1429W} 
  Wende S., Reiners A., Ludwig H.-G., 2009, A\&A, 508, 1429 
 \bibitem[Wu et al.(2011)Wu et al.]{2011RAA....11..924W} 
  Wu Y., et al., 2011, RAA, 11, 924 
\end{thebibliography}
\end{document}